\documentclass[a4paper,fleqn,usenatbib]{mnras}

\usepackage{newtxtext,newtxmath}

\usepackage[T1]{fontenc}
\usepackage{ae,aecompl}

\usepackage{graphicx}	
\usepackage{url}
\usepackage{amsmath}	

\usepackage{amssymb}	
\usepackage{subfigure}
\usepackage{makecell}
 \usepackage{color}

\title[DIROL]{Data--driven Image Restoration with Option--driven Learning for Big and Small Astronomical Image Datasets}

\author[Peng Jia et al.]{
Peng Jia,$^{1,2,3}$\thanks{robinmartin20@gmail.com}
Ruiyu Ning,$^{1}$, Ruiqi Sun$^1$, Xiaoshan Yang$^{1}$ and Dongmei Cai$^{1}$\\
$^{1}$College of Physics and Optoelectronics, Taiyuan University of Technology, Taiyuan, 030024, China\\
$^{2}$Key Laboratory of Advanced Transducers and Intelligent Control Systems, Ministry of Education and Shanxi Province, \\Taiyuan University of Technology, Taiyuan, 030024, China\\
$^{3}$Department of Physics, Durham University, South Road, Durham, DH1 3LE, UK \\
}

\date{Accepted XXX. Received YYY; in original form ZZZ}

\pubyear{2015}

\begin{document}
\label{firstpage}
\pagerange{\pageref{firstpage}--\pageref{lastpage}}
\maketitle

\begin{abstract}
Image restoration methods are commonly used to improve the quality of astronomical images. In recent years, developments of deep neural networks and increments of the number of astronomical images have evoked a lot of data--driven image restoration methods. However, most of these methods belong to supervised learning algorithms, which require paired images either from real observations or simulated data as training set. For some applications, it is hard to get enough paired images from real observations and simulated images are quite different from real observed ones. In this paper, we propose a new data--driven image restoration method based on generative adversarial networks with option--driven learning. Our method uses several high resolution images as references and applies different learning strategies when the number of reference images is different. For sky surveys with variable observation conditions, our method can obtain very stable image restoration results, regardless of the number of reference images.
\end{abstract}

\begin{keywords}
techniques: image processing -- methods: numerical
\end{keywords}



\section{Introduction}
\label{sec:1}
The quality of astronomical images are seriously limited by different effects such as: photon--electronic noises from the detector, point spread functions of the imaging system and sky background noises. To further improve the quality of astronomical images, different image restoration methods are proposed for images of different astronomical targets observed in different wavelengths \citep{narayan1986maximum, starck1995multiresolution, bertero2000image,  esch2004image, kuwamura2008image, la2012airy, jia2014parallel, xu2020solar, fetick2020blind}.\\

Image restoration methods can be classified into two different categories according to their prior conditions: image restoration methods based on modelling of the point spread function (PSF) or image restoration methods based on extraction of image features (IF). PSF-based image restoration methods firstly model PSFs according to telemetry data of telescopes and environmental data \citep{martin2016psf, fetick2019physics, beltramo2019prime, Jia2020c, fusco2020reconstruction} or images of point sources \citep{jia2017blind, sun2020improving}. Then these methods could obtain restored images with prior PSFs through either deconvolution \citep{starck2002deconvolution} or myopic--deconvolution algorithms \citep{conan1998myopic, fusco1999myopic, mugnier2001myopic}. However, for images of extended targets (such as intergalactic medium, nebulae or the Sun), obtaining point sources is quite hard \citep{long2019point} and limited point sources in these images are often polluted by light from extended medium. Besides, ground--based sky survey telescopes normally can not provide effective telemetry data for PSF modelling. Under these circumstances, IF--based image restoration methods would be a possible way to improve image qualities.\\

IF--based image restoration methods are successors of image based regularization conditions in blind deconvolution algorithms \citep{carasso2001direct}. Classical regularization conditions such as the total variation condition \citep{chan1998total}, the normalized sparsity measure condition \citep{krishnan2011blind}, dictionary--based conditions \citep{namba1998wavelet}, noise and signal properties \citep{prato2013convergent} and similarities between continuous frames of images \citep{schulz1993multiframe} are all IF conditions that are drawn from objective world by human experts. However, designing a general IF condition that is suitable for different kinds of astronomical images is very hard.\\

In recent years, a lot of optical and infrared astronomical images have been collected and released to the public \citep{burstein1994beijing, scoville2007cosmos, zhao2007china, grogin2011candels, liu2014new, alam2015eleventh, ma2018first}. In the future, there would be huge volume of optical astronomical images obtained by the Vera Rubin Observatory \citep{ivezic2019lsst}, the Euclid Satellite \citep{laureijs2010euclid}, future ground based large telescopes \citep{johns2006giant, gilmozzi2007european, sanders2013thirty, cui2016chinese} and other survey projects \citep{zhao2011probing, benitez2014j, liu2020sitian}. With this unprecedentedly large volume of astronomical data, human beings would be able to obtain millions to billions images of different extended astronomical targets. With the help of statistical pattern recognition \citep{webb2003statistical} and deep learning \citep{goodfellow2016deep}, IF conditions of different extended astronomical targets could be obtained and corresponding IF--based image restoration methods will also become possible.\\

A preliminary IF-based image restoration method was proposed by \citet{jia2019solar} for solar image restoration. Because solar images of the same wavelength are representations of the same physical process, we find that texture features in solar images of the same wavelength satisfies the same probability distribution \citep{huang2019perception}. Then with several high resolution solar images as references, the CycleGAN \citep{zhu2017unpaired} can restore any frames of solar images of the same wavelength. Because there is only one sun for us to observe, we could always obtain a lot of high resolution solar images of different wavelengths either with adaptive optic systems \citep{zhang2017solar} or through some post--processing methods \citep{li2014parallel, xiang2016high} as references. Because it is possible to get large number of effective reference images, solar image restoration can be viewed as data-driven image restoration task for big astronomical image datasets. With adequate algorithms, we could always obtain effective restored images.\\

For night time observations, although the total number of extended astronomical targets is large, the number of images of some targets with a specific kind is small. For example, there are only 100 images of planetary nebulae in [N II] wavelength in the Atlas of monochromatic images of planetary nebulae \citep{weidmann2016atlas}. Besides, many of these extended targets have relatively low surface brightness. These targets require very long exposure time to get images with enough signal to noise ratio and some of them can only be obtained at the end of a sky survey project through stacking observed images of several epochs. For these astronomical targets, high resolution images that can be used as references are rare.\\

Deep neural networks (DNNs) used for data--driven image restoration are generative models, such as the Generative Adversarial Networks \citep{wang2017generative} or Variational Autoencoders \citep{kingma2019introduction}. The generative models learn map functions between two probability distributions represented by data in the training set. Because the probability distribution is discretely sampled by several images, the number and varieties of images will affect their representative abilities. Finally the quality of restored images will become unstable, which makes these algorithms unsuitable for scientific research. Although data--driven image restoration algorithms have good performance when the number of reference images is big, its performance will become bad when the number of reference images is small. To reduce requirements of data volume in data--driven image restoration methods, we propose a data--driven image restoration method with option--driven learning (DIROL). The DIROL is a GAN with new structure which could adjust constraint conditions according to the number of reference images. We will introduce the DIROL in Section \ref{sec:2} and compare the performance of the DIROL with other image restoration methods in Section \ref{sec:3}. In Section \ref{sec:4}, we will make our conclusions and anticipate our future works.\\

\section{The Data--Driven image restoration method with option--driven learning}
\label{sec:2}
Data--Driven image restoration methods obtain IF conditions through statistical learning. Because DNNs have very strong representation ability, IF conditions are often embedded into DNNs. After training, DNNs can generate restored images from blurred images. Generative Adversarial Networks (GAN) and Variational Autoencoders (VAE) are commonly used generative models. The VAE is an autoencoder that regularizes its encoding distribution during the training stage to guarantee that the latent space of the VAE can generate reliable restored images from blurred images. However, because the latent space between blurred images and restored images is not regular, images generated by VAEs are not stable \citep{sami2019comparative}. The GAN can generate images with better quality. Through modifications of the structure of GANs, adaptions of different loss functions or different training strategies, different types of GANs are proposed, such as the CGAN \citep{mirza2014conditional}, the Wasserstein GAN \citep{arjovsky2017wasserstein} and the CycleGAN \citep{zhu2017unpaired}. Although the performance of GANs in image restoration tasks has been increased,  lack of reference images is still a problem.\\

The GalaxyGAN is the first successful GAN used for astronomical image restoration \citep{schawinski2017generative}. The GalaxyGAN uses pairs of images to train the GAN. Because these images are generated through numerical simulation with real datasets, there are always enough paired images for the GalaxyGAN. Thanks to its generalization property, the GalaxyGAN can restore blurred images after training. However, as we discussed in Section \ref{sec:1}, high resolution images in the paired images are hard to obtain in some applications.\\

To reduce requirements of large amount of high resolution images, we propose the CycleGAN to restore Solar images \citep{jia2019solar} and a similar method is also proposed for restoration of images with realistic blur \citep{zhang2020deblurring}. The CycleGAN does not need paired images as a training set. It can restore blurred images with several high resolution images as references. The CycleGAN has been used to restore astronomical images of different types. In real applications, we find that the instability of GAN will bring serious problems to restored images including: model collapse, loss of information and generation of additional components. Insufficient and improper constraints obtained from data are main factors that magnify shortcomings of GANs  \citep{wang2019evolutionary}. Heuristically, we could strengthen constraints of the probability space represented by reference images and limit influence of the discriminator on the entire network when reference images are not adequate. Based on this philosophy, we propose the DIROL for image restoration.\\

\subsection{Main structure of the DIROL}
\label{sec:MainStructure}
\begin{figure*}
\centering
\includegraphics[height=7cm]{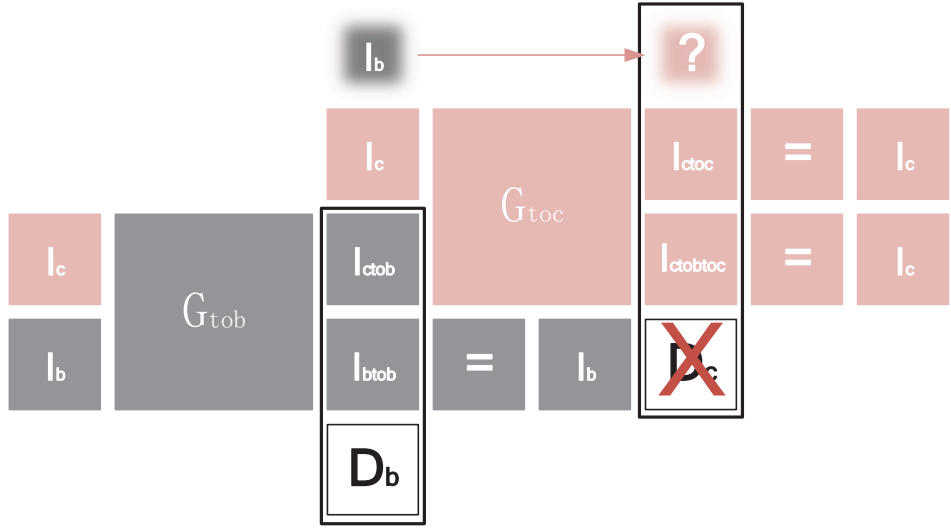}
\caption{The structure of the DIROL. It has an Optional--Driven Learning strategy: active learning and passive learning for different number of reference images. For the active learning, we use losses calculated by the $D_{c}$ between $I_c$ and  $I_{btoc}$ to train $G_{toc}$. For the passive learning, we disable the $D_{c}$ and use the L1 losses between $I_c$ and $I_{ctobtoc}$ or $I_c$ and $I_{ctoc}$ to train $G_{toc}$. In the diagram, we use gray to indicate that the component is currently in a degraded distribution domain, and pink to indicate that the component is currently in a high-resolution distribution domain.}
\label{fig:odl}
\end{figure*}

The DIROL is a self-supervised learning algorithm. We select high resolution images $I_c$ as reference images and $I_b$ are images that are used to restore. After training, the DIROL directly restores $I_b$ and can also restore other images with similar blur level. The main structure of the DIROL is shown in figure \ref{fig:odl}. It has two generators ($G$) and two discriminators ($D$). $G_{tob}$ generates blurred images from input images. $G_{toc}$ generates high resolution images from input images. $D$ discriminates categories of images in its corresponding box. $I_c$ and $I_b$ stand for high resolution images and blurred images.  $I_{ctob}$ stands for blurred images generated from high resolution images and $I_{btob}$ stands for blurred images generated from blurred images. $I_{ctoc}$ stands for high resolution images generated from high resolution images. $I_{ctobtoc}$ stands for high resolution images generated from $I_{ctob}$.\\

In ordinary GANs, images are generated by generators according to losses obtained by discriminators. Through integrating of several generators and discriminators, the solution space can be strictly restricted. Under this philosophy, we have proposed three major improvements for the DIROL:\\

1. Introduction of the self--mapping loss to the DIROL. Conventional conditional generative neural networks often use single-path mapping to generate images. We propose the self--mapping loss, which could generate two--path mapping to further increase stability of the neural network. We will discuss it in Section \ref{selfmapping}.\\

2. Introduction of a pre--training stage to the DIROL. In GANs, generators and discriminators are often coupled. There would be some risks that GANs would trap in local minimal values. We propose to use pre--training strategy to increase the performance of the DIROL. We will discuss this strategy in Section \ref{ParaInit}.\\

3. Introduction of the Option--Driven Learning (ODL) strategy to the DIROL. Considering the number of reference images, we propose the ODL in real applications. For reference images of large volume, we apply the active learning method and passive learning method for reference images of small volume. We will discuss it in Section \ref{ODLLearning}.\\

With these improvements, we design the DIROL as a GAN with cycle consistency structure as shown in figure \ref{fig:odl}. The DIROL is initialized with the pre--training method discussed in Section \ref{ParaInit}. Then according to number and properties of reference images, we use the ODL to train the DIROL with hyper--parameters defined in Table \ref{table:hyperparameter setting} and loss function defined in Equation \ref{eq:total},\\

\begin{table}
	\caption{Hyper-parameters of the DIROL in this paper.}
	\begin{tabular}{ccc}
		\hline
		Parameters  & Values \\
		\hline
		Pre-training Iterations &  5000\\
		Training Iterations & 30000\\
		Batch Size & 1\\
		Learning Rate & 0.0001\\
		Weight of Self Loss & 20\\
		Weight of Cycle Loss & 20\\
		Weight of Similarity Loss & 0.1\\
		\hline 						
	\end{tabular}	
	\label{table:hyperparameter setting}
\end{table}

\begin{equation}
\begin{aligned}
\label{eq:total}
L(G,D_b)  = L_{odl} + \lambda_i L_{similarity}
\end{aligned}
\end{equation}
where $L_{similarity}$ is the similarity loss function defined in equation \ref{eq:similarity}. $L_{odl}$ is ODL loss which will be further discussed in Section \ref{ODLLearning}. We use the Adam optimizer \cite{kingma2014adam} as the parameter update strategy and the details of the network structure can be found in the appendix \ref{appendix}.\\
\begin{equation}
\begin{aligned}
\label{eq:similarity}
L_{similarity}(G_{toc},G_{tob})  = & E_{b\sim{p_{data}(b)}}[\parallel G_{toc}(b)-b\parallel_1] \\
                  + & E_{c\sim{p_{data}(c)}}[\parallel G_{tob}(c)-c \parallel_1]
\end{aligned}
\end{equation}

\subsection{Self--mapping Loss for Image Restoration}
\label{selfmapping}
\begin{figure*}
\centering
\subfigure[two-path mapping]{
\begin{minipage}[t]{0.5\textwidth}
\centering
\includegraphics[height=7cm]{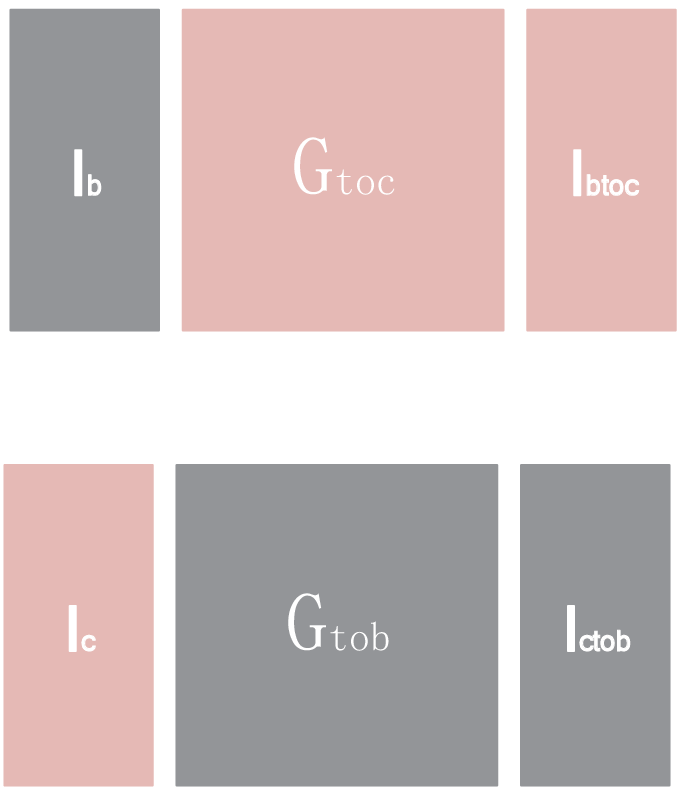}
\label{fig:self-mapping_a}
\end{minipage}
}%
\subfigure[one-path mapping]{
\begin{minipage}[t]{0.5\textwidth}
\centering
\includegraphics[height=7cm]{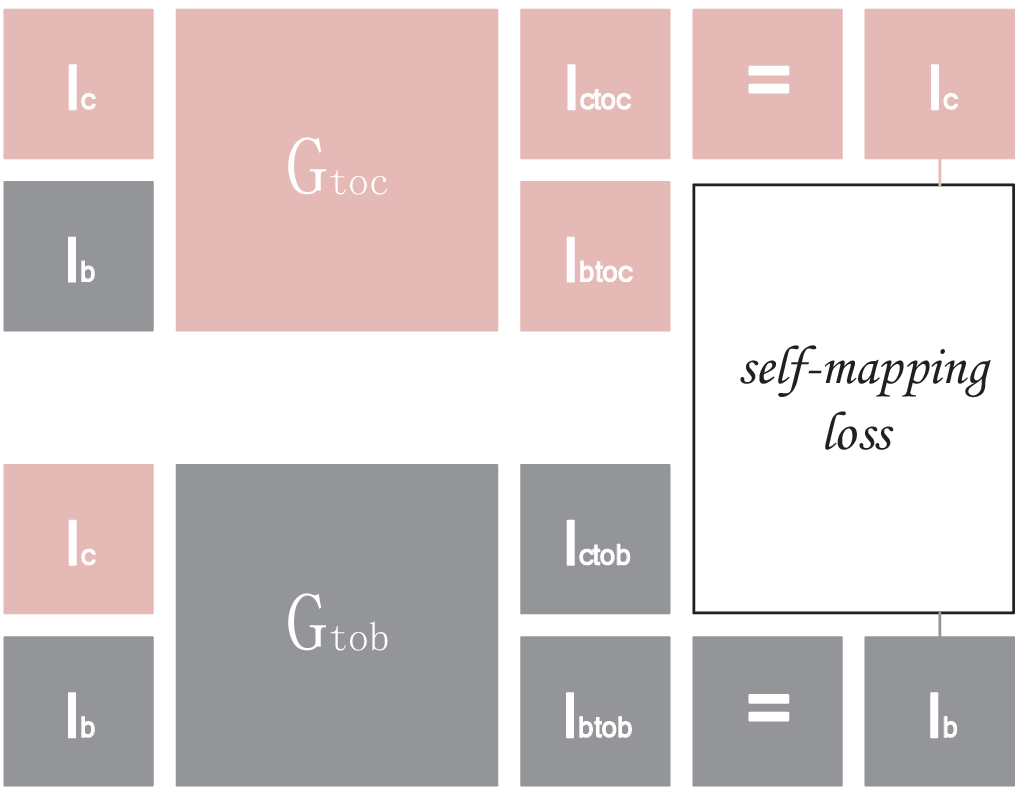}
\label{fig:self-mapping_b}
\end{minipage}
}%
\centering
\caption{Diagrams of one-path mapping and two-path mapping. The self-mapping loss in (a) provides a regular term for the ${G_{toc}}$ and ${G_{tob}}$, while the one-path mapping in (b) makes the generation network lack of sufficient constraints. In the diagram, we use gray to indicate that the component is currently in a degraded distribution domain, and pink to indicate that the component is currently in a high-resolution distribution domain.}
\label{fig:self-mapping}
\end{figure*}

The one--path mapping is a commonly used structure in conditional generative neural networks. The structure of the one--path mapping is shown in figure \ref{fig:self-mapping_a}. The one--path mapping will directly map blurred images to high resolution images and vice versa. Through connecting several one--path mappings, the performance of generative neural networks could be improved. For example, the CycleGAN introduces a cycle consistency loss between images generated by two one--path mappings to further improve its performance.\\

However, the one--path mapping is weak in restricting solution space of generators, because discriminators will only compare probability distributions represented by images. In this paper, we further introduce self--mapping loss as shown in figure \ref{fig:self-mapping_b}. With the self--mapping loss, generators will use high resolution images and blurred images at the same time as references. Then we will compare high resolution images generated from high resolution images with original high resolution images as loss function. Meanwhile, we will also compare blurred images generated from blurred images with original blurred images as loss function too. The definition of self--mapping loss is defined in equation \ref{eq:self-mapping loss}. Because the self--mapping loss is $L_1$ norm, it will better restrict the solution space, comparing with discriminator that are defined in probability space. With self--mapping loss, we can generate two--path mapping to further increase the performance of the GAN. \\

\begin{equation}
\begin{aligned}
\label{eq:self-mapping loss}
L_{self}(G_{toc},G_{tob})  = & E_{c\sim{p_{data}(c)}}[\parallel G_{toc}(c)-c\parallel_1] \\
                           + & E_{b\sim{p_{data}(b)}}[\parallel G_{tob}(b)-b\parallel_1]
\end{aligned}
\end{equation}
\subsection{Parameter Initialization}
\label{ParaInit}
\begin{figure*}
\centering
\subfigure[Generator]{
\begin{minipage}[t]{0.5\textwidth}
\centering
\includegraphics[width=7cm]{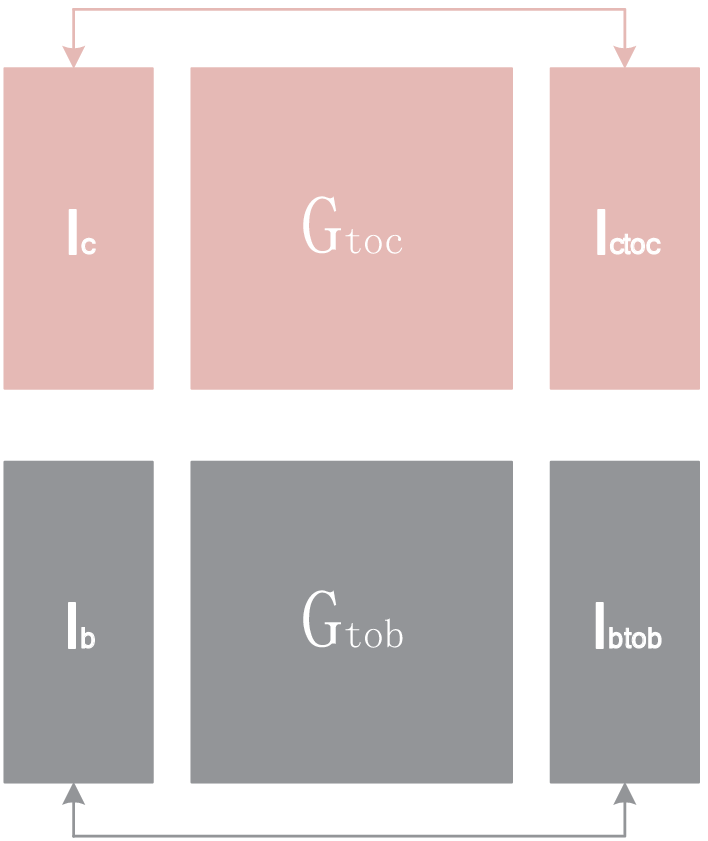}
\end{minipage}
}%
\subfigure[Discriminator]{
\begin{minipage}[t]{0.5\textwidth}
\centering
\includegraphics[width=8cm,height=3.8cm]{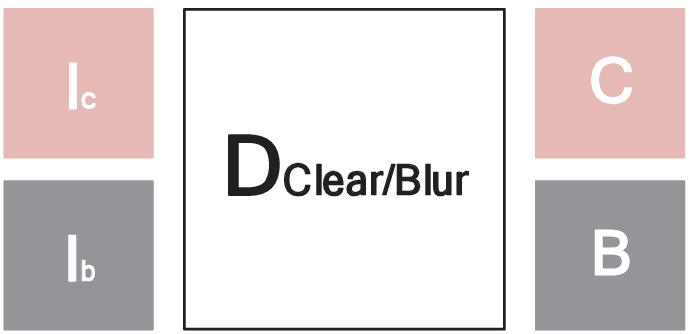}
\end{minipage}
}%
\centering
\caption{Pre-training process. The generator G in (a) is pre--trained by images with self-mapping loss. The discriminator D in (b) is pre-trained through the prior category information of high resolution images and blurred images. In the diagram, we use gray to indicate that the component is currently in a degraded distribution domain, and pink to indicate that the component is currently in a high-resolution distribution domain.}
\label{fig:pretraining}
\end{figure*}

Proper parameter initialization is beneficial to training of neural networks. Particularly for GANs, whose generators and discriminators are coupled, proper initializations are essential. In this paper, we initialize the generator and the discriminator as shown in figure \ref{fig:pretraining} with loss function defined in equation \ref{eq:pre-training loss}. The generator is pre--trained with images according to self--mapping loss. The discriminator is pre--trained to classify high resolution images and blurred images. According to our experience in designing loss functions in CycleGAN, we use the least-squares loss function instead of log likelihood function to improve the performance of the network \citep{mao2017least}.\\

\begin{equation}
\begin{aligned}
\label{eq:pre-training loss}
L_{pretraining}(G,D)  = & L_{self} \\
                  -& E_{b\sim{p_{data}(b)}}[log(D_b(b))+ log(1-D_c(b))] \\
                  -& E_{c\sim{p_{data}(c)}}[log(D_c(c))+ log(1-D_b(c))] 
\end{aligned}
\end{equation}

\subsection{Option-Driven Learning strategy}
\label{ODLLearning}
The DIROL adapts the successful cycle structure that is proposed in the CycleGAN and uses cycle loss defined in equation \ref{eq:cycle loss} as one of its loss functions. Besides, the DIROL has two modes in real applications: passive learning and active learning. The active learning uses losses from $D_{c}$ between $I_c$ and ${I_{btoc}}$ or $D_{c}$ between ${I_c}$ and ${I_{ctobtoc}}$  to train $G_{toc}$  when the volume of data is big, as shown in equation \ref{eq:odl-a}.\\
\begin{equation}
\begin{aligned}
\label{eq:cycle loss}
L_{cycle}(G_{toc},G_{tob})  = & E_{c\sim{p_{data}(c)}}[\parallel G_{toc}(G_{tob}(c))-c\parallel_1] \\
                           + & E_{b\sim{p_{data}(b)}}[\parallel G_{tob}(G_{toc}(b))-b\parallel_1]
\end{aligned}
\end{equation}

\begin{equation}
\begin{aligned}
\label{eq:odl-a}
L_{odl-a}(G,D) = &\lambda_s L_{self} + \lambda_c L_{cycle} \\
                  +& E_{b\sim{p_{data}(b)}}[log(D_b(b))+ log(1-D_c(G_{toc}(b)))] \\
                  +& E_{c\sim{p_{data}(c)}}[log(D_c(c)) + log(1-D_b(G_{tob}(c)))]\\
\end{aligned}
\end{equation}

When the number or varieties of reference images reduce, the performance of generators will quickly drop down. Under this circumstance, we use the passive learning strategy. The passive learning disables the $D_{c}$ to better restrict qualities of restored images and the loss function in passive learning is defined in equation \ref{eq:odl-p}.\\
\begin{equation}
\begin{aligned}
\label{eq:odl-p}
L_{odl-p}(G_{toc},G_{tob},D_b) = &\lambda_s L_{self} + \lambda_c L_{cycle} \\
                  +& E_{b\sim{p_{data}(b)}}[log(D_b(b)] \\
                  +& E_{c\sim{p_{data}(c)}}[log(1-D_b(G_{tob}(c)))]\\
\end{aligned}
\end{equation}

The cycle loss defined in equation \ref{eq:cycle loss} provides basic conditions for $G_{toc}$ to generate ${I_{btoc}}$ from $I_b$. The self--mapping loss between ${I_{ctoc}}$ and ${I_{c}}$ provides a regularization condition for ${G_{toc}}$. The loss from the discriminator and that from the self--mapping are options for the DIROL in reconstruction of $I_{b}$. With the ODL strategy and the two--path mapping, the DIROL can restore blurred images effectively.\\

\section{Performance of the DIROL}
\label{sec:3}
In this section, we will compare the performance of DIROL with that of two other methods: the CycleGAN and the GalaxyGAN. The GalaxyGAN is a supervised image restoration method \citet{schawinski2017generative}, which is trained by simulated images of different blur levels and restore images based on its generalization ability. The GalaxyGAN can restore blurred images within its blur level. When the observation condition changes, its performance would be affected. The CycleGAN \citep{jia2019solar} is an unsupervised image restoration method, which only requires high resolution as references and does not sensitive to blur levels. However, the CycleGAN uses probability distribution obtained by the discriminator as loss function. When reference images are not adequate or the number of reference images is too small, the CycleGAN would generate fake structure. We will show shortcomings of these methods with simulated data and show that DIROL has some advantages over these two methods in this section.\\  

\subsection{Datasets and Evaluation Protocols}
To test the performance of our algorithm, we use a dataset that consists of 4550 galaxies from the SDSS Data Release 12 as reference images \citep{stoughton2002sloan}. We generate simulated degraded images with the method provided in \citet{schawinski2017generative}. We firstly stretch the greyscale of g, r and i band images with asinh transformation. Then we convolve these images by Gaussian function with different full width half magnitude (FWHM) to generate blurred images. Through adding different levels of noises to blurred images, we can obtain degraded images with different noise levels. Parameters used in this paper are shown in table \ref{table:experimental design}.\\ 

\begin{table}
	\caption{Parameters used in this paper to generate simulated images.}
	\begin{tabular}{ccc}
		\hline
		FWHM  & Noise-Level \\
		\hline
		1.4 &  2\\
		1.8 & 5 (used in section \ref{MoreDiscussion})\\
		2.5 & 10\\
		\hline 						
	\end{tabular}	
	\label{table:experimental design}
\end{table}

We set two scenarios in this part: image restoration with small amount of data as references and image restoration with large amount of data as references. For small data scenario, the amount of data in the training set is $20 \times 2$ frames (20 frames of high resolution images and corresponding blurred images). For big data scenario, the amount of data in the training set is $2000 \times 2$ frames (2000 frames of high resolution images and corresponding blurred images).\\

We will evaluate the quality of images obtained by the GalaxyGAN, the CycleGAN, the DIROL with active learning (DIROL--A) and the DIROL with passive learning (DIROL--P) in this paper. Because seeing conditions and noise levels will change during real observations and the GalaxyGAN is trained by paired data, we will use images with FWHM of 1.4 and noise level of 2 to train the neural network and images with any blur levels and noise level of 10 as test set to reflect real conditions. Because other image restoration methods are unsupervised learning algorithms, we will directly use images with any blur levels and noise level to test the performance of them.\\

Two evaluation metrics are used in this paper: the peak signal to noise ratio (PSNR) defined in Scikit-Image \citep{van2014scikit} and the Fréchet Inception Distance (FID) defined by \citet{salimans2016improved}. The PSNR is defined in equation \ref{eq:psnr},
\begin{equation}
\begin{aligned}
\label{eq:psnr}
PSNR = 20\cdot log_{10} (MAX_1) - 10\cdot log_{10} (MSE), \\
MSE = \frac{1}{mn} \sum_{i=0}^{m-1} \sum_{j=0}^{n-1} [I(i,j) - K(i,j)]^2.
\end{aligned}
\end{equation}
where $I$ and $K$ are original and restored images with size of $m\times n$, $MAX_1$ is the maximal greyscale in restored images. The PSNR is a classical image quality metric. From the definition, we can find that the PSNR can directly reflect similarity between restored images and blurred images. Images with larger PSNR are better. However, the PSNR requires the original image as reference, which makes it not adequate for real image quality evaluation.\\

The FID is a classical statistical image quality metric that are used to evaluate the performance of GANs. The FID is evaluated with features extracted from images by a pre--trained Inception V3 model \citep{szegedy2016rethinking}. It is defined in equation \ref{eq:fid},
\begin{equation}
\begin{aligned}
\label{eq:fid}
FID = \| \mu_1 - \mu_2 \| + Tr(C_1 + C_2 - 2 \times (C_1 \times C_2)^2)
\end{aligned}
\end{equation}
where $\mu_1$ and $\mu_2$ stand for feature--wise mean of features extracted from real observed images and generated images, $C_1$ and $C_2$ stand for covariance matrix of features from real observed images and generated images. $Tr$ stands for trace calculation between two matrix. The FID measures the distance between distributions represented by high resolution images and blurred images. The FID is different from PSNR. It is consistent with human visual judgment and does not need original high resolution images, which make it adequate for real image quality evaluation. Restored images with smaller FID values are more similar to high resolution images, which means they have better image quality.\\

\subsection{Image Restoration Results with Small Datasets}
\label{LearnFromSmall}
\begin{figure*}
\centering
\includegraphics[height=7cm]{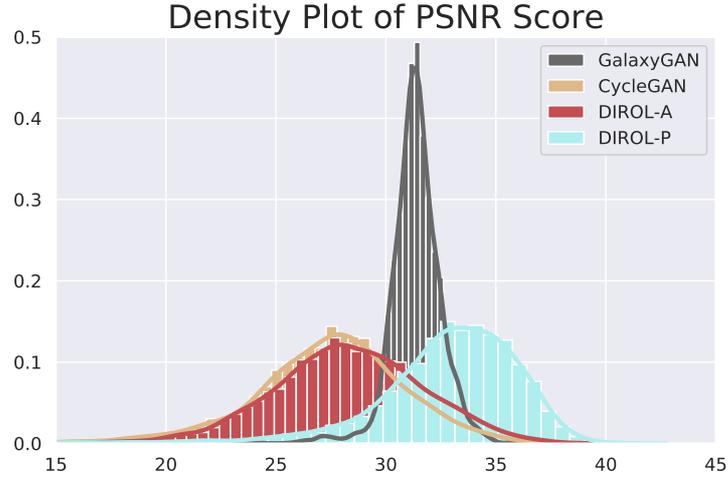}
\caption{PSNR statistical histogram of restored images, when DNNs are trained by reference images with small number. The average value of PSNR for different methods are 32.80 for DIROL--P, 31.37 for GalaxyGAN, 28.30 for DIROL--A and 27.49 for CycleGAN.}
\label{fig:SmallPSNR}
\end{figure*}

Firstly, we consider image restoration in small data scenario. According to our discussions in Section \ref{ODLLearning}, when the number of reference images is small, passive learning strategy is required to avoid uncertainty caused by the discriminator. PSNR statistical histogram is shown in figure \ref{fig:SmallPSNR}. As we can find in this figure that the DIROL--P has the best PSNR. However, it should be noted that PSNR of images restored by the GalaxyGAN is more concentrated, which indicates that the supervised learning is more stable.\\

The FID of images obtained by these methods is shown in table \ref{SmallFID1}. Because the FID is a statistical metric, we use 2000 restored images obtained by different set of training to evaluate our method. We can find that the FID for images obtained by the DIROL--P is very small and that obtained by the GalaxyGAN is large. It indicates us that the DIROL has the best performance. However, it should be noted that there are some artificial structures generated by GANs. As shown in figure \ref{fig:SmallResult}, all GANs will generate artificial structures. When the number of reference images is small, over--fitting will change contents of images. Considering both the FID and the PSNR, for images of small dataset, the DIROL--P has the best performance.\\
 
\begin{table*}
    \begin{center}
	\caption{This table shows the reconstruction effect of training the network with small amount of data, which is evaluated using FID score.}.
    \label{SmallFID1}
	\begin{tabular}{ccccc}
	\hline
	 FID  &2.5-10&1.8-10&1.4-10&mean\\
	\hline
	GalaxyGAN  &3.9420&3.9558&3.9609&3.9529\\
      CycleGAN    &0.9956&0.9301&1.3364&1.0874\\
	DIROL--A    &0.7312&0.8004&0.4397&0.6571\\
	DIROL--P   &0.1238&0.8038&0.5435&0.4904\\
	\hline
	\end{tabular}
    \end{center}
\end{table*}

\subsection{Image Restoration Results with Big Datasets}
\label{LearnFromLarge}
\begin{figure*}
\centering
\includegraphics[height=7cm]{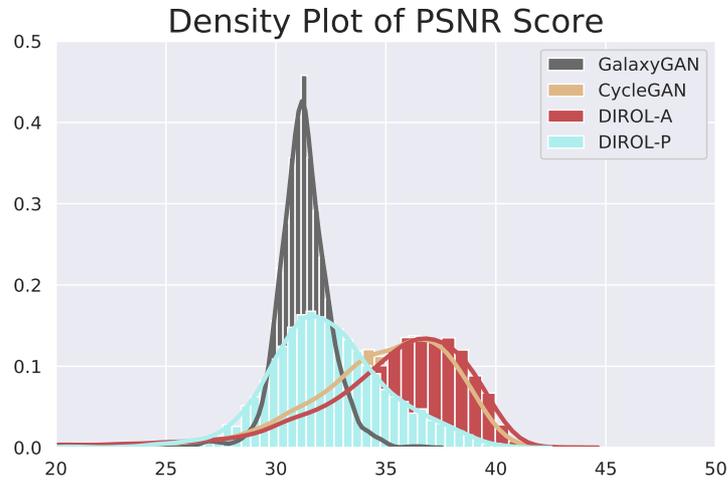}
\caption{PSNR statistical histogram of restored images, when DNNs are trained by reference images with big number. The average value of PSNR for different methods are 34.92 for DIROL--A, 34.74 for CycleGAN, 32.36 for DIROL--P  and 31.26 for GalaxyGAN.}
\label{fig:LargePSNR}
\end{figure*}

Then we consider image restoration problem in big data scenario. For this scenario, DIROL will choose a different strategy. With more reference images,  additional constraints provided by the discriminator can provide better constraint condition. Therefore, we need to use the DIROL--A method. The PSNR histogram of restored images by different methods is shown in figure \ref{fig:LargePSNR}. We can find that results obtained by DIROL--A has the best quality.\\

We further use the FID to evaluate the quality of restored images. The results are shown in table \ref{tab:LargeFID}. We can find that the FID for images obtained by the DIROL--A is very small and that obtained by the GalaxyGAN is large. However, thanks to increment of reference images, the over--fitting problem is not serious. As shown in figure \ref{fig:LargeResult}, there are no artificial structure in restored images. Considering the FID and PSNR, the DIROL--A has better performance.\\

\begin{table*}
    \begin{center}
	\caption{This table shows the reconstruction effect of training the network with large amount of data, which is evaluated using FID score.}
    \label{tab:LargeFID}
	\begin{tabular}{ccccc}
	\hline
	 FID  &2.5-10&1.8-10&1.4-10&mean\\
	\hline
	GalaxyGAN  &1.3622&1.3680&1.3712&1.3671\\
      CycleGAN    &0.1710&0.1080&0.0475&0.1088\\
	DIROL--A      &0.0390&0.0746&0.1555&0.0897\\
	DIROL--P   &0.6211&1.0021&0.5721&0.7318\\
	\hline
	\end{tabular}
    \end{center}
\end{table*}

\subsection{More Discussions}
\label{MoreDiscussion}
\begin{figure*}
\centering
\includegraphics[height=7cm]{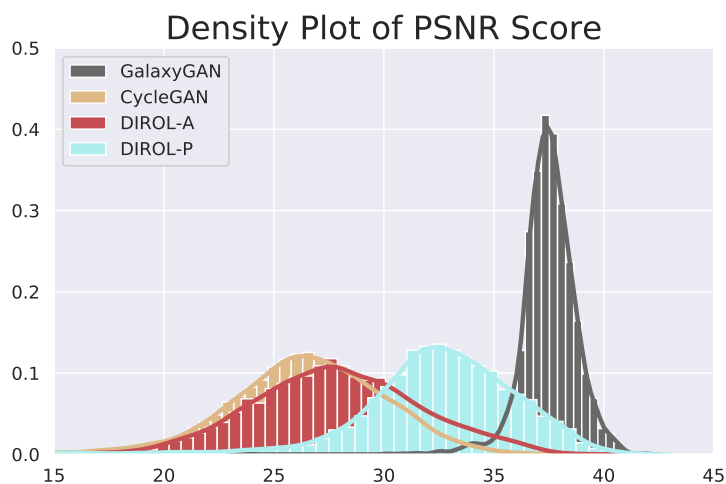}
\caption{PSNR statistical histogram of images restored by DNNs trained by small datasets, when DNNs are tested by images with different blur levels and noise level of 5. The average value of PSNR for different methods are 37.50 for GalaxyGAN , 32.57 for DIROL--P, 27.86 for DIROL--A and 26.56 for CycleGAN.}
\label{fig:MoreSmallPSNR}
\end{figure*}

\begin{figure*}
\centering
\includegraphics[height=7cm]{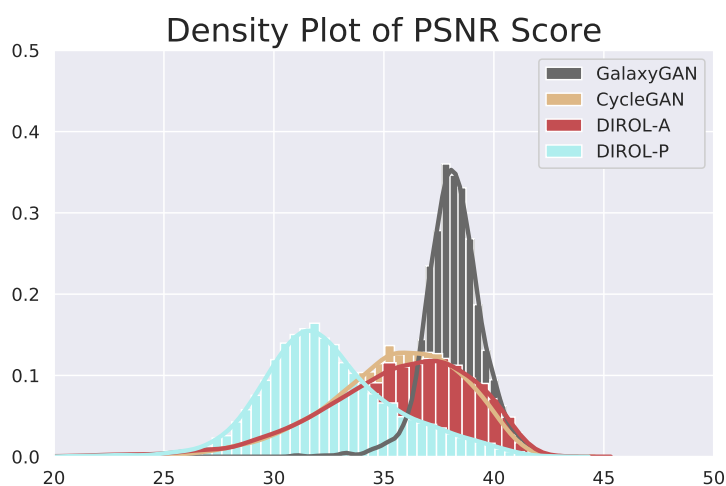}
\caption{PSNR statistical histogram of images restored by DNNs trained by big datasets, when DNNs are tested by images with different blur levels and noise level of 5. The average value of PSNR for different methods are 38.16 for GalaxyGAN , 35.44 for DIROL--A, 35.44 for CycleGAN and 32.49 for DIROL--P.}
\label{fig:MoreLargePSNR}
\end{figure*}

In previous sections, we have shown that the DIROL has good performance when the observation conditions change or there are different number of reference images. There are some possibilities that if the observation condition does not change much, image restoration based on supervised learning can obtain good results. As shown in figure \ref{fig:MoreSmallPSNR} and figure \ref{fig:MoreLargePSNR}, we use data with a noise level of 2 for training and data with a noise level of 5 for testing. As shown in these figures, GalaxyGAN can get the highest PSNR score compared to all generative model based image restoration methods. However, considering variations of blur type, level and noise level, it is hard to obtain training set that can represent all levels of blur and noise. Therefore, the DIROL is better than other supervised image restoration method as the performance of it is independent of simulation conditions of the training data. \\

\section{Conclusions and future works}
\label{sec:4}
We have presented a simple but effective GAN based image restoration method -- DIROL in this paper. The DIROL uses different strategies when the number of reference images is different. Tested with simulated data, we can find that the DIROL can obtain stable image restoration results. Although the DIROL shows some advantages than other methods, instability of GANs, such as model collapse or over--fitting are still inevitable. In our future works, we will design additional physical--based IFs and integrate PSF model to further improve the performance of image restoration methods.

\section*{Acknowledgements}
This work is supported by National Natural Science Foundation of China (NSFC) (11503018, 61805173), the Joint Research Fund in Astronomy (U1631133) under cooperative agreement between the NSFC and Chinese Academy of Sciences (CAS). Authors acknowledge the French National Research Agency (ANR) to support this work through the ANR APPLY (grant ANR-19-CE31-0011) coordinated by B. Neichel. This work is also supported by Shanxi Province Science Foundation for Youths (201901D211081), Research and Development Program of Shanxi (201903D121161), Research Project Supported by Shanxi Scholarship Council of China (HGKY2019039), the Scientific and Technological Innovation Programs of Higher Education Institutions in Shanxi (2019L0225). PJ would like to thank Professor Hui Liu and Professor Kaifan Ji from Yunnan Astronomical Observatory, Professor Li Ji and Professor Jiangtao Li from Purple Mountain Observatory, Dr. Renato Dupke from Brazil National Observatory for their helpful suggestions.\\
Data Availability Statements: the code in this paper can be downloaded from aojp.lamost.org and it will be released in PaperData Repository Powered by China-VO with a DOI number.\\


\bibliographystyle{mnras}
\bibliography{DIROL} 

\begin{thebibliography}{}
\makeatletter
\relax
\def\mn@urlcharsother{\let\do\@makeother \do\$\do\&\do\#\do\^\do\_\do\%\do\~}
\def\mn@doi{\begingroup\mn@urlcharsother \@ifnextchar [ {\mn@doi@}
  {\mn@doi@[]}}
\def\mn@doi@[#1]#2{\def\@tempa{#1}\ifx\@tempa\@empty \href
  {http://dx.doi.org/#2} {doi:#2}\else \href {http://dx.doi.org/#2} {#1}\fi
  \endgroup}
\def\mn@eprint#1#2{\mn@eprint@#1:#2::\@nil}
\def\mn@eprint@arXiv#1{\href {http://arxiv.org/abs/#1} {{\tt arXiv:#1}}}
\def\mn@eprint@dblp#1{\href {http://dblp.uni-trier.de/rec/bibtex/#1.xml}
  {dblp:#1}}
\def\mn@eprint@#1:#2:#3:#4\@nil{\def\@tempa {#1}\def\@tempb {#2}\def\@tempc
  {#3}\ifx \@tempc \@empty \let \@tempc \@tempb \let \@tempb \@tempa \fi \ifx
  \@tempb \@empty \def\@tempb {arXiv}\fi \@ifundefined
  {mn@eprint@\@tempb}{\@tempb:\@tempc}{\expandafter \expandafter \csname
  mn@eprint@\@tempb\endcsname \expandafter{\@tempc}}}

\bibitem[\protect\citeauthoryear{Alam et~al.,}{Alam
  et~al.}{2015}]{alam2015eleventh}
Alam S.,  et~al., 2015, The Astrophysical Journal Supplement Series, 219, 12

\bibitem[\protect\citeauthoryear{Arjovsky, Chintala  \& Bottou}{Arjovsky
  et~al.}{2017}]{arjovsky2017wasserstein}
Arjovsky M.,  Chintala S.,   Bottou L.,  2017, arXiv preprint arXiv:1701.07875

\bibitem[\protect\citeauthoryear{Beltramo-Martin, Correia, Ragland, Jolissaint,
  Neichel, Fusco  \& Wizinowich}{Beltramo-Martin
  et~al.}{2019}]{beltramo2019prime}
Beltramo-Martin O.,  Correia C.,  Ragland S.,  Jolissaint L.,  Neichel B.,
  Fusco T.,   Wizinowich P.,  2019, Monthly Notices of the Royal Astronomical
  Society, 487, 5450

\bibitem[\protect\citeauthoryear{Benitez et~al.,}{Benitez
  et~al.}{2014}]{benitez2014j}
Benitez N.,  et~al., 2014, arXiv preprint arXiv:1403.5237

\bibitem[\protect\citeauthoryear{Bertero \& Boccacci}{Bertero \&
  Boccacci}{2000}]{bertero2000image}
Bertero M.,  Boccacci P.,  2000, Astronomy and Astrophysics Supplement Series,
  147, 323

\bibitem[\protect\citeauthoryear{Burstein et~al.,}{Burstein
  et~al.}{1994}]{burstein1994beijing}
Burstein D.,  et~al., 1994, AAS, 185, 41

\bibitem[\protect\citeauthoryear{Carasso}{Carasso}{2001}]{carasso2001direct}
Carasso A.~S.,  2001, SIAM Journal on Applied Mathematics, 61, 1980

\bibitem[\protect\citeauthoryear{Chan \& Wong}{Chan \&
  Wong}{1998}]{chan1998total}
Chan T.~F.,  Wong C.-K.,  1998, IEEE transactions on Image Processing, 7, 370

\bibitem[\protect\citeauthoryear{Conan, Mugnier, Fusco, Michau  \&
  Rousset}{Conan et~al.}{1998}]{conan1998myopic}
Conan J.-M.,  Mugnier L.~M.,  Fusco T.,  Michau V.,   Rousset G.,  1998,
  Applied Optics, 37, 4614

\bibitem[\protect\citeauthoryear{Cui \& Zhu}{Cui \& Zhu}{2016}]{cui2016chinese}
Cui X.-q.,  Zhu Y.-t.,  2016, in Ground-based and Airborne Telescopes VI. p.
  990607

\bibitem[\protect\citeauthoryear{Esch, Connors, Karovska  \& van Dyk}{Esch
  et~al.}{2004}]{esch2004image}
Esch D.~N.,  Connors A.,  Karovska M.,   van Dyk D.~A.,  2004, The
  Astrophysical Journal, 610, 1213

\bibitem[\protect\citeauthoryear{F{\'e}tick et~al.,}{F{\'e}tick
  et~al.}{2019}]{fetick2019physics}
F{\'e}tick R.,  et~al., 2019, Astronomy \& Astrophysics, 628, A99

\bibitem[\protect\citeauthoryear{F{\'e}tick, Mugnier, Fusco  \&
  Neichel}{F{\'e}tick et~al.}{2020}]{fetick2020blind}
F{\'e}tick R.~J.,  Mugnier L.,  Fusco T.,   Neichel B.,  2020, Monthly Notices
  of the Royal Astronomical Society, 496, 4209

\bibitem[\protect\citeauthoryear{Fusco, V{\'e}ran, Conan  \& Mugnier}{Fusco
  et~al.}{1999}]{fusco1999myopic}
Fusco T.,  V{\'e}ran J.-P.,  Conan J.-M.,   Mugnier L.,  1999, Astronomy and
  Astrophysics Supplement Series, 134, 193

\bibitem[\protect\citeauthoryear{Fusco et~al.,}{Fusco
  et~al.}{2020}]{fusco2020reconstruction}
Fusco T.,  et~al., 2020, Astronomy \& Astrophysics, 635, A208

\bibitem[\protect\citeauthoryear{Gilmozzi \& Spyromilio}{Gilmozzi \&
  Spyromilio}{2007}]{gilmozzi2007european}
Gilmozzi R.,  Spyromilio J.,  2007, The Messenger, 127, 3

\bibitem[\protect\citeauthoryear{Goodfellow, Bengio, Courville  \&
  Bengio}{Goodfellow et~al.}{2016}]{goodfellow2016deep}
Goodfellow I.,  Bengio Y.,  Courville A.,   Bengio Y.,  2016, Deep learning.
 Vol. 1, MIT press Cambridge

\bibitem[\protect\citeauthoryear{Grogin et~al.,}{Grogin
  et~al.}{2011}]{grogin2011candels}
Grogin N.~A.,  et~al., 2011, The Astrophysical Journal Supplement Series, 197,
  35

\bibitem[\protect\citeauthoryear{Huang, Jia, Cai  \& Cai}{Huang
  et~al.}{2019}]{huang2019perception}
Huang Y.,  Jia P.,  Cai D.,   Cai B.,  2019, Solar Physics, 294, 133

\bibitem[\protect\citeauthoryear{Ivezi{\'c} et~al.,}{Ivezi{\'c}
  et~al.}{2019}]{ivezic2019lsst}
Ivezi{\'c} {\v{Z}}.,  et~al., 2019, The Astrophysical Journal, 873, 111

\bibitem[\protect\citeauthoryear{Jia, Cai  \& Wang}{Jia
  et~al.}{2014}]{jia2014parallel}
Jia P.,  Cai D.,   Wang D.,  2014, Experimental Astronomy, 38, 41

\bibitem[\protect\citeauthoryear{Jia, Sun, Wang, Cai  \& Liu}{Jia
  et~al.}{2017}]{jia2017blind}
Jia P.,  Sun R.,  Wang W.,  Cai D.,   Liu H.,  2017, Monthly Notices of the
  Royal Astronomical Society, 470, 1950

\bibitem[\protect\citeauthoryear{Jia, Huang, Cai  \& Cai}{Jia
  et~al.}{2019}]{jia2019solar}
Jia P.,  Huang Y.,  Cai B.,   Cai D.,  2019, The Astrophysical Journal Letters,
  881, L30

\bibitem[\protect\citeauthoryear{{Jia}, {Wu}, {Yi}, {Cai}  \& {Cai}}{{Jia}
  et~al.}{2020}]{Jia2020c}
{Jia} P.,  {Wu} X.,  {Yi} H.,  {Cai} B.,   {Cai} D.,  2020, \mn@doi [\aj]
  {10.3847/1538-3881/ab7b79}, \href
  {https://ui.adsabs.harvard.edu/abs/2020AJ....159..183J} {159, 183}

\bibitem[\protect\citeauthoryear{Johns}{Johns}{2006}]{johns2006giant}
Johns M.,  2006, in Ground-based and Airborne Telescopes. p. 626729

\bibitem[\protect\citeauthoryear{Kingma \& Ba}{Kingma \&
  Ba}{2014}]{kingma2014adam}
Kingma D.~P.,  Ba J.,  2014, arXiv preprint arXiv:1412.6980

\bibitem[\protect\citeauthoryear{Kingma \& Welling}{Kingma \&
  Welling}{2019}]{kingma2019introduction}
Kingma D.~P.,  Welling M.,  2019, arXiv preprint arXiv:1906.02691

\bibitem[\protect\citeauthoryear{Krishnan, Tay  \& Fergus}{Krishnan
  et~al.}{2011}]{krishnan2011blind}
Krishnan D.,  Tay T.,   Fergus R.,  2011, in CVPR 2011. pp 233--240

\bibitem[\protect\citeauthoryear{Kuwamura, Tsumuraya, Miura  \& Baba}{Kuwamura
  et~al.}{2008}]{kuwamura2008image}
Kuwamura S.,  Tsumuraya F.,  Miura N.,   Baba N.,  2008, Publications of the
  Astronomical Society of the Pacific, 120, 348

\bibitem[\protect\citeauthoryear{La~Camera, Carbillet, Olivieri, Boccacci  \&
  Bertero}{La~Camera et~al.}{2012}]{la2012airy}
La~Camera A.,  Carbillet M.,  Olivieri C.,  Boccacci P.,   Bertero M.,  2012,
  in Optical and Infrared Interferometry III. p. 84453E

\bibitem[\protect\citeauthoryear{Laureijs, Duvet, Sanz, Gondoin, Lumb,
  Oosterbroek  \& Criado}{Laureijs et~al.}{2010}]{laureijs2010euclid}
Laureijs R.~J.,  Duvet L.,  Sanz I.~E.,  Gondoin P.,  Lumb D.~H.,  Oosterbroek
  T.,   Criado G.~S.,  2010, in Space Telescopes and Instrumentation 2010:
  Optical, Infrared, and Millimeter Wave. p. 77311H

\bibitem[\protect\citeauthoryear{Li, Wang, Xiang, Zheng, Liu, Deng  \& Ji}{Li
  et~al.}{2014}]{li2014parallel}
Li X.-B.,  Wang F.,  Xiang Y.~Y.,  Zheng Y.~F.,  Liu Y.~B.,  Deng H.,   Ji
  K.~F.,  2014, Journal of The Korean Astronomical Society, 47, 43

\bibitem[\protect\citeauthoryear{Liu et~al.,}{Liu et~al.}{2014}]{liu2014new}
Liu Z.,  et~al., 2014, Research in Astronomy and Astrophysics, 14, 705

\bibitem[\protect\citeauthoryear{Liu, Soria, Wu, Wu  \& Shang}{Liu
  et~al.}{2020}]{liu2020sitian}
Liu J.,  Soria R.,  Wu X.-F.,  Wu H.,   Shang Z.,  2020, arXiv preprint
  arXiv:2006.01844

\bibitem[\protect\citeauthoryear{Long, Soubo, Weiping, Feng  \& Jun}{Long
  et~al.}{2019}]{long2019point}
Long M.,  Soubo Y.,  Weiping N.,  Feng X.,   Jun Y.,  2019, The Astrophysical
  Journal, 888, 20

\bibitem[\protect\citeauthoryear{Ma et~al.,}{Ma et~al.}{2018}]{ma2018first}
Ma B.,  et~al., 2018, Monthly Notices of the Royal Astronomical Society, 479,
  111

\bibitem[\protect\citeauthoryear{Mao, Li, Xie, Lau, Wang  \& Paul~Smolley}{Mao
  et~al.}{2017}]{mao2017least}
Mao X.,  Li Q.,  Xie H.,  Lau R.~Y.,  Wang Z.,   Paul~Smolley S.,  2017, in
  Proceedings of the IEEE international conference on computer vision. pp
  2794--2802

\bibitem[\protect\citeauthoryear{Martin et~al.,}{Martin
  et~al.}{2016}]{martin2016psf}
Martin O.,  et~al., 2016, in Adaptive Optics Systems V. p. 99091Q

\bibitem[\protect\citeauthoryear{Mirza \& Osindero}{Mirza \&
  Osindero}{2014}]{mirza2014conditional}
Mirza M.,  Osindero S.,  2014, arXiv preprint arXiv:1411.1784

\bibitem[\protect\citeauthoryear{Mugnier, Robert, Conan, Michau  \&
  Salem}{Mugnier et~al.}{2001}]{mugnier2001myopic}
Mugnier L.~M.,  Robert C.,  Conan J.-M.,  Michau V.,   Salem S.,  2001, JOSA A,
  18, 862

\bibitem[\protect\citeauthoryear{Namba \& Ishida}{Namba \&
  Ishida}{1998}]{namba1998wavelet}
Namba M.,  Ishida Y.,  1998, Signal processing, 68, 119

\bibitem[\protect\citeauthoryear{Narayan \& Nityananda}{Narayan \&
  Nityananda}{1986}]{narayan1986maximum}
Narayan R.,  Nityananda R.,  1986, Annual review of astronomy and astrophysics,
  24, 127

\bibitem[\protect\citeauthoryear{Prato, La~Camera, Bonettini  \& Bertero}{Prato
  et~al.}{2013}]{prato2013convergent}
Prato M.,  La~Camera A.,  Bonettini S.,   Bertero M.,  2013, Inverse Problems,
  29, 065017

\bibitem[\protect\citeauthoryear{Salimans, Goodfellow, Zaremba, Cheung, Radford
   \& Chen}{Salimans et~al.}{2016}]{salimans2016improved}
Salimans T.,  Goodfellow I.,  Zaremba W.,  Cheung V.,  Radford A.,   Chen X.,
  2016, in Advances in neural information processing systems. pp 2234--2242

\bibitem[\protect\citeauthoryear{Sami \& Mobin}{Sami \&
  Mobin}{2019}]{sami2019comparative}
Sami M.,  Mobin I.,  2019, in 2019 International Conference of Artificial
  Intelligence and Information Technology (ICAIIT). pp~1--5

\bibitem[\protect\citeauthoryear{Sanders}{Sanders}{2013}]{sanders2013thirty}
Sanders G.~H.,  2013, Journal of Astrophysics and Astronomy, 34, 81

\bibitem[\protect\citeauthoryear{Schawinski, Zhang, Zhang, Fowler  \&
  Santhanam}{Schawinski et~al.}{2017}]{schawinski2017generative}
Schawinski K.,  Zhang C.,  Zhang H.,  Fowler L.,   Santhanam G.~K.,  2017,
  Monthly Notices of the Royal Astronomical Society: Letters, 467, L110

\bibitem[\protect\citeauthoryear{Schulz}{Schulz}{1993}]{schulz1993multiframe}
Schulz T.~J.,  1993, JOSA A, 10, 1064

\bibitem[\protect\citeauthoryear{Scoville et~al.,}{Scoville
  et~al.}{2007}]{scoville2007cosmos}
Scoville N.,  et~al., 2007, The Astrophysical Journal Supplement Series, 172,
  38

\bibitem[\protect\citeauthoryear{Starck, Murtagii  \& Bijaoui}{Starck
  et~al.}{1995}]{starck1995multiresolution}
Starck J.-L.,  Murtagii F.,   Bijaoui A.,  1995, Graphical models and image
  processing, 57, 420

\bibitem[\protect\citeauthoryear{Starck, Pantin  \& Murtagh}{Starck
  et~al.}{2002}]{starck2002deconvolution}
Starck J.-L.,  Pantin E.,   Murtagh F.,  2002, Publications of the Astronomical
  Society of the Pacific, 114, 1051

\bibitem[\protect\citeauthoryear{Stoughton et~al.,}{Stoughton
  et~al.}{2002}]{stoughton2002sloan}
Stoughton C.,  et~al., 2002, The Astronomical Journal, 123, 485

\bibitem[\protect\citeauthoryear{Sun, Yu, Jia  \& Zhao}{Sun
  et~al.}{2020}]{sun2020improving}
Sun R.,  Yu S.,  Jia P.,   Zhao C.,  2020, Monthly Notices of the Royal
  Astronomical Society, 497, 4000

\bibitem[\protect\citeauthoryear{Szegedy, Vanhoucke, Ioffe, Shlens  \&
  Wojna}{Szegedy et~al.}{2016}]{szegedy2016rethinking}
Szegedy C.,  Vanhoucke V.,  Ioffe S.,  Shlens J.,   Wojna Z.,  2016, in
  Proceedings of the IEEE conference on computer vision and pattern
  recognition. pp 2818--2826

\bibitem[\protect\citeauthoryear{Van~der Walt, Sch{\"o}nberger, Nunez-Iglesias,
  Boulogne, Warner, Yager, Gouillart  \& Yu}{Van~der Walt
  et~al.}{2014}]{van2014scikit}
Van~der Walt S.,  Sch{\"o}nberger J.~L.,  Nunez-Iglesias J.,  Boulogne F.,
  Warner J.~D.,  Yager N.,  Gouillart E.,   Yu T.,  2014, PeerJ, 2, e453

\bibitem[\protect\citeauthoryear{Wang, Gou, Duan, Lin, Zheng  \& Wang}{Wang
  et~al.}{2017}]{wang2017generative}
Wang K.,  Gou C.,  Duan Y.,  Lin Y.,  Zheng X.,   Wang F.-Y.,  2017, IEEE/CAA
  Journal of Automatica Sinica, 4, 588

\bibitem[\protect\citeauthoryear{Wang, Xu, Yao  \& Tao}{Wang
  et~al.}{2019}]{wang2019evolutionary}
Wang C.,  Xu C.,  Yao X.,   Tao D.,  2019, IEEE Transactions on Evolutionary
  Computation, 23, 921

\bibitem[\protect\citeauthoryear{Webb}{Webb}{2003}]{webb2003statistical}
Webb A.~R.,  2003, Statistical pattern recognition.
John Wiley \& Sons

\bibitem[\protect\citeauthoryear{Weidmann, Schmidt, Valdarenas, Ahumada, Volpe
  \& Mudrik}{Weidmann et~al.}{2016}]{weidmann2016atlas}
Weidmann W.,  Schmidt E.,  Valdarenas R.~V.,  Ahumada J.,  Volpe M.,   Mudrik
  A.,  2016, Astronomy \& Astrophysics, 592, A103

\bibitem[\protect\citeauthoryear{Xiang, Liu  \& Jin}{Xiang
  et~al.}{2016}]{xiang2016high}
Xiang Y.-y.,  Liu Z.,   Jin Z.-y.,  2016, New Astronomy, 49, 8

\bibitem[\protect\citeauthoryear{Xu, Sun, Yan  \& Zhang}{Xu
  et~al.}{2020}]{xu2020solar}
Xu L.,  Sun W.,  Yan Y.,   Zhang W.,  2020, arXiv preprint arXiv:2001.03850

\bibitem[\protect\citeauthoryear{Zhang, Guo  \& Rao}{Zhang
  et~al.}{2017}]{zhang2017solar}
Zhang L.,  Guo Y.,   Rao C.,  2017, Optics Express, 25, 4356

\bibitem[\protect\citeauthoryear{Zhang, Luo, Zhong, Ma, Stenger, Liu  \&
  Li}{Zhang et~al.}{2020}]{zhang2020deblurring}
Zhang K.,  Luo W.,  Zhong Y.,  Ma L.,  Stenger B.,  Liu W.,   Li H.,  2020, in
  Proceedings of the IEEE/CVF Conference on Computer Vision and Pattern
  Recognition. pp 2737--2746

\bibitem[\protect\citeauthoryear{Zhao, Yao  \& Lu}{Zhao
  et~al.}{2007}]{zhao2007china}
Zhao H.,  Yao J.,   Lu H.,  2007, Proceedings of the International Astronomical
  Union, 3, 565

\bibitem[\protect\citeauthoryear{Zhao, Zhan, Wang, Fan  \& Zhang}{Zhao
  et~al.}{2011}]{zhao2011probing}
Zhao G.-B.,  Zhan H.,  Wang L.,  Fan Z.,   Zhang X.,  2011, Publications of the
  Astronomical Society of the Pacific, 123, 725

\bibitem[\protect\citeauthoryear{Zhu, Park, Isola  \& Efros}{Zhu
  et~al.}{2017}]{zhu2017unpaired}
Zhu J.-Y.,  Park T.,  Isola P.,   Efros A.~A.,  2017, in Proceedings of the
  IEEE international conference on computer vision. pp 2223--2232

\makeatother
\end{thebibliography}



\appendix
\section{The Structure of DNNs in the DIROL}
\label{appendix}
\begin{table}
\caption{The structure of Generator in the DIROL.}
\begin{tabular}{ccc}
\hline
Generator & kernel size/stride & output\\
\hline
Conv2d & 7*7*1& 212*212*32\\
IN & & 212*212*32\\
Relu & & 212*212*32\\
Conv2d & 3*3*2& 106*106*64\\
IN & & 106*106*64\\
Relu & & 106*106*64\\
Conv2d & 3*3*2& 53*53*128\\
IN & & 53*53*128\\
Relu & & 53*53*128\\
ResidualBlock& & 53*53*128\\
ResidualBlock& & 53*53*128\\
ResidualBlock& & 53*53*128\\
ResidualBlock& & 53*53*128\\
ResidualBlock& & 53*53*128\\
ResidualBlock& & 53*53*128\\
ConvT2d & 3*3*2& 106*106*64\\
IN & & 106*106*64\\
Relu & & 106*106*64\\
ConvT2d & 3*3*2& 212*212*32\\
IN & & 212*212*32\\
Relu & & 212*212*32\\
Conv2d & 7*7*1& 212*212*1\\
\hline 
\end{tabular} 
\label{table:Generator}
\end{table}
\begin{table}
\caption{The structure of Discriminator in the DIROL.}
\begin{tabular}{ccc}
\hline
Discriminator & kernel size/stride & output\\
\hline
Conv2d & 4*4*2& 106*106*32\\
LeakyRelu &   & 106*106*32\\
Conv2d & 4*4*2& 53*53*64\\
IN    &       & 53*53*64\\
LeakyRelu &    & 53*53*64\\
Conv2d & 4*4*1& 52*52*128\\
IN     &      & 52*52*128\\
LeakyRelu &    & 52*52*128\\
Conv2d & 4*4*1& 51*51*1\\
Sigmoid\\
\hline 
\end{tabular} 
\label{table:Discriminator}
\end{table}
\begin{table}
\caption{The structure of the ResidualBlock.}
\begin{tabular}{ccc}
\hline
ResidualBlock & kernel size/stride\\
\hline
Conv2d & 3*3*1\\
IN     &      \\
Relu   &      \\
Conv2d & 3*3*1\\
IN     &      \\
\hline 
\end{tabular} 
\label{table:residual}
\end{table}

\section{Blurred and Restored Images obtained by different methods.}
\begin{figure*}
\centering
\includegraphics[height=20cm]{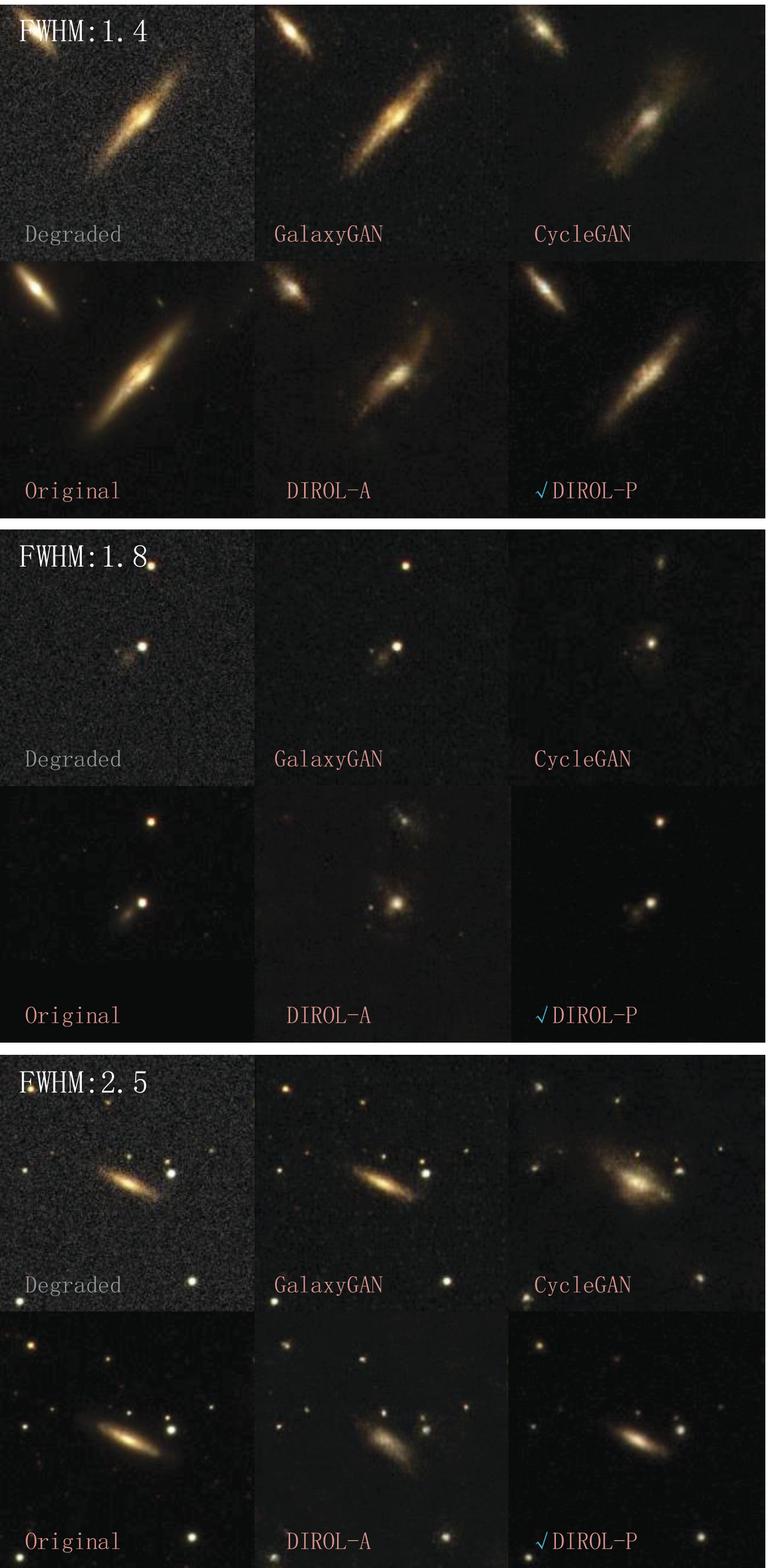}
\caption{Images restored by different methods with a small amount of data.}
\label{fig:SmallResult}
\end{figure*}

\begin{figure*}
\centering
\includegraphics[height=20cm]{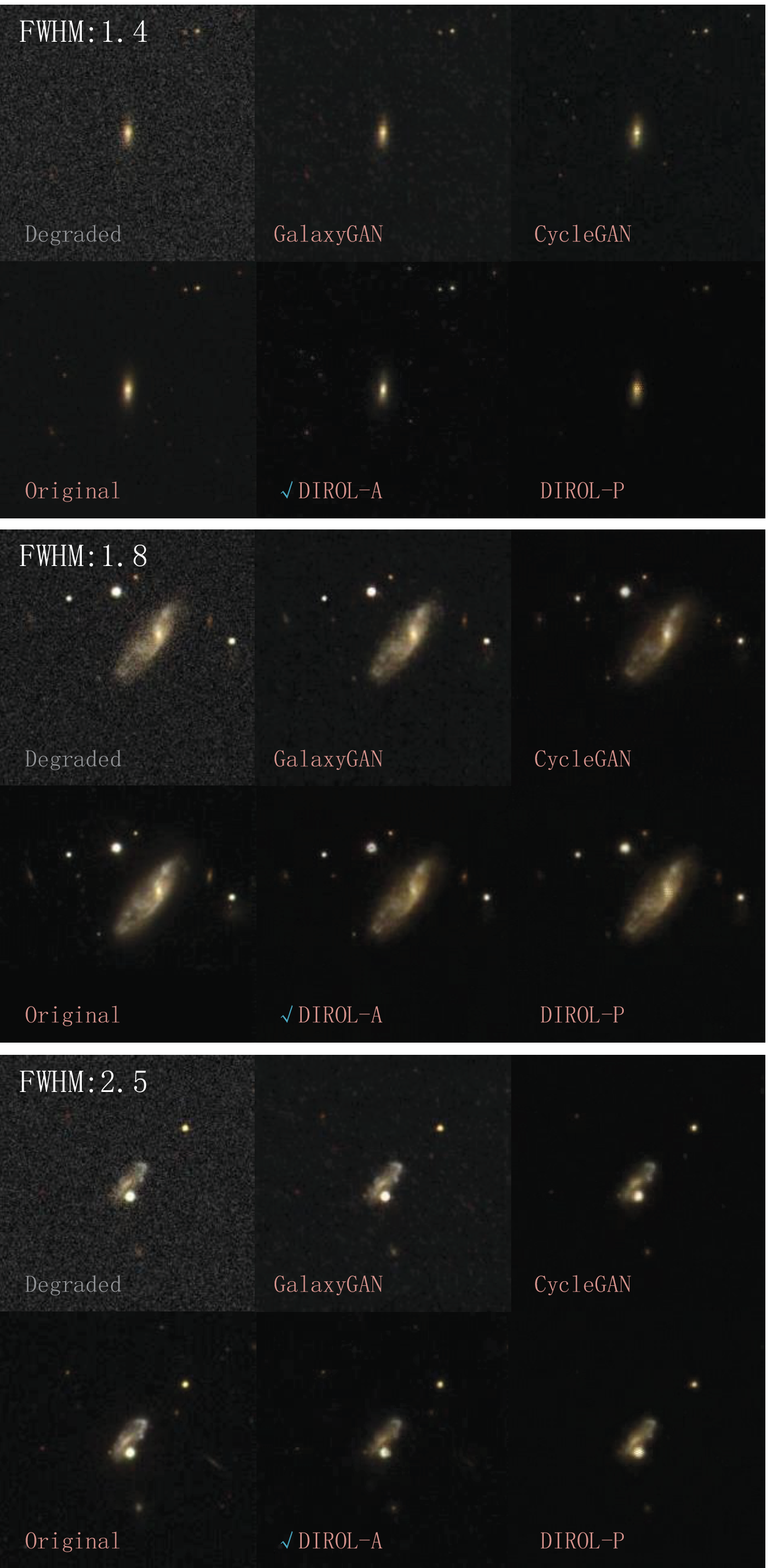}
\caption{Images restored by different methods with a big amount of data.}
\label{fig:LargeResult}
\end{figure*}


\bsp	
\label{lastpage}
\end{document}